\documentclass[aps,superscriptaddress,showpacs, twocolumn]{revtex4}

\usepackage{makeidx}
\usepackage{epsfig}
\usepackage{amsfonts}
\usepackage{amsmath}
\usepackage{amssymb}
\usepackage{graphicx}
\usepackage{epstopdf}
\usepackage{float}

\newcommand{\bM}{{\bf M}}                                   
\newcommand{\bm}{{\bf m}}                                   
\newcommand{\Ms}{M_{\rm s}}                                 

\newcommand{\Hx }{{    H}_{\rm x}}                          

\newcommand{\Hz }{{    H}_{\rm z}}                          

\newcommand{\ex}{\hat{\bf e}_{\rm x}}                       
\newcommand{\ez}{\hat{\bf e}_{\rm z}}                       

\newcommand{\bHeff}{{\bf H}_{\rm eff}}                      

\newcommand{\GLL }{{\bf \Gamma}_{\rm LL }}                  
\newcommand{\GGD }{{\bf \Gamma}_{\rm GD }}                  
\newcommand{\GST }{{\bf \Gamma}_{\rm ST }}                  

\newcommand{\ep  }{{\bf m}_{\rm p}}                         
\newcommand{\vIs }{\vec{\cal I}_{\rm s  }}                  
\newcommand{\Is  }{    {\cal I}_{\rm s  }}                  
\newcommand{\Ic  }{    {\cal I}_{\rm c  }}                  
\newcommand{\etax}{\eta_{\rm x}}                            
\newcommand{\etaz}{\eta_{\rm z}}                            

\newcommand{\IDC }{\overline{\cal I}_{\rm s  }}             
\newcommand{\IDCc}{\overline{\cal I}_{\rm c  }}             
\newcommand{\IDCo}{\overline{\cal I}_{\rm opt}}             

\newcommand{\IAC }{\widetilde{\cal I}_{\rm s  }}            
\newcommand{\IACo}{\widetilde{\cal I}_{\rm opt}}            
\newcommand{\IACc}{\widetilde{\cal I}_{\rm c  }}            

\newcommand{\RAC}{R}                                        
\newcommand{\RDC}{R}                                        

\newcommand{\tpl}{{t_{\rm pl}}}                             
\newcommand{\tAC}{{t_{\rm AC}}}                             
\newcommand{\tDC}{{t_{\rm DC}}}                             

\newcommand{\Ei}{{E_{\rm 0}}}                               
\newcommand{\Eb}{{E_{\rm b}}}                               
\newcommand{\Ec}{{E_{\rm c}}}                               
\newcommand{\Eeq}{E_{\rm eq}}                               

\newcommand{\VDC }{\overline{V}        }                    
\newcommand{\VAC }{\widetilde{V}        }                   
\newcommand{\WAC }{\widetilde{W}        }                   

\newcommand{\dEAC}{\dot{E}}                                 
\newcommand{\dEDC}{\dot{E}}                                 

\newcommand{\omo}{\omega_{\rm opt}}                         

\newcommand{\REF}{Ref.\ }
\newcommand{\REFS}{Refs.\ }
\newcommand{\FIG}{Fig.\ }

\newcommand{\EQ}{Eqn.\ }
\newcommand{\EQS}{Eqs.\ }

\begin{document}

\title{Optimization of spin-torque switching using AC and DC pulses}

\author{Tom Dunn$^1$, and  Alex Kamenev$^{1,2}$ \\
 $^1$Department of Physics, University of Minnesota, Minneapolis, Minnesota 55455, USA. \\
 $^2$Fine Theoretical Physics Institute, University of Minnesota, Minneapolis, Minnesota 55455, USA.}

\begin{abstract}
We explore spin-torque induced magnetic reversal in magnetic tunnel junctions using combined AC and DC spin-current pulses. We calculate the optimal pulse times and  current strengths for both AC and DC pulses as well as the optimal AC signal frequency, needed to minimize the Joule heat lost during the switching process. The results of this optimization are compared against numeric simulations. Finally we show how this optimization leads to different dynamic regimes, where switching is optimized by either a purely AC or DC spin-current, or a combination AC/DC spin-current, depending on the anisotropy energies and the spin-current polarization.
\end{abstract}

\maketitle

\section{Introduction}
\label{sec:introduction}
The question of spin-torque (ST) efficiency in switching the free layer in magnetic tunnel junctions (MTJ) has been of great interest in recent years \cite{Nikinov10,Fricke12,Carpentieri12,Seo12} due to spintronic memory's potential as a universal non-volatile memory element \cite{Yuasa07,Matsunaga09}. For MTJ with pinned and free layer easy-axis parallel the most efficient prescription has been shown both theoretically and experimentally to be a DC current pulse roughly twice the critical current \cite{Bedau10,Dunn10}. Recent experimental research, however, has shown MTJ with a second pinned layer, polarized perpendicular to the free layer easy-axis, along the free layer easy-plane axis, can produce switching much faster and more efficiently than in similar co-linear devices \cite{Kent04,Rowlands11}. Furthermore, previous theoretical work by the authors of this paper has shown AC spin-current pulses in MTJ with strong free layer easy-plane anisotropy and with both parallel and perpendicularly polarized pinned layers can improve the efficiency further by inducing a resonant response in the free layer \cite{Dunn12_JAP}.

For MTJs with weak easy-plane anisotropy (such as those used in Ref. \cite{Amiri11,Worledge11,Zhao12}) this purely AC method becomes less effective. This is a result of the magnetization spending more time at large azimuthal angles through the switching process, where the ST from the perpendicular pinned layer is weaker. Conversely, as the magnetization spends more time at large azimuthal angles the strength of the ST from the parallel pinned layer ST gets larger for DC currents. In fact, for some cases the strengths of the AC and DC ST may intersect allowing the AC ST to dominate for low energy orbits and the DC ST for high energy orbits. This tradeoff suggests an alternative means of magnetic switching in MTJs with weak easy-plane anisotropy. Instead of using a purely AC or DC spin-current, an AC pulse may be used to push the magnetization to a higher energy state, where a DC spin-current can then be used to switch the magnetization the rest of the way. Such an AC/DC current pulse strategy is considered experimentally in Ref. \cite{Cui08} and theoretically using micromagnetic simulations in Ref. \cite{Carpentieri12} and shown to markedly improve the efficiency of the switching process.

\begin{figure}[H]
  \begin{centering}
  \includegraphics[width=8.6cm]{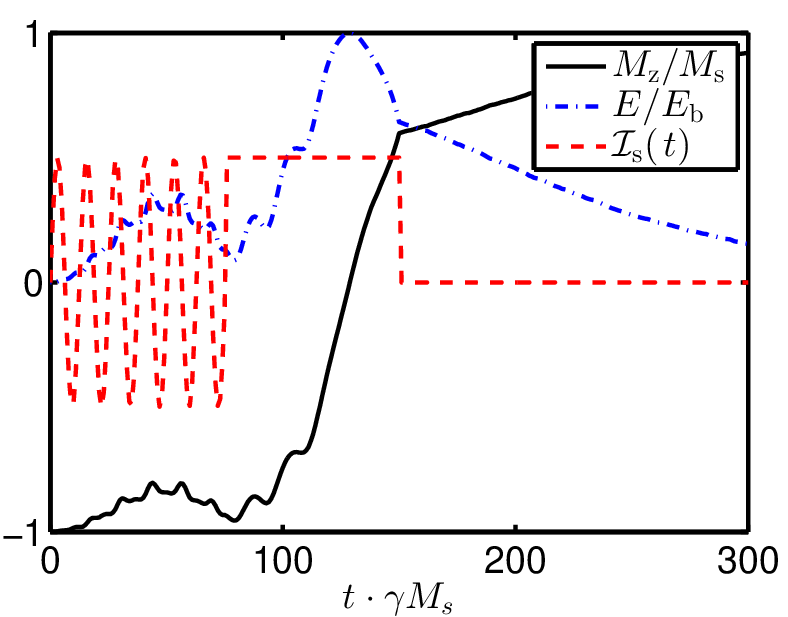}
  \par\end{centering}
  \caption{(Color online) Simulated switching trajectory calculated via numeric integration of the LLS equation; $M_\mathrm{z}$ (black,solid) and energy $E$ (blue,dot-dashed) along with spin-current in arbitrary units (red,dashed). Here   $\Hx = 0.5\,\Ms$, $\Hz = 0$, $\alpha = 0.015$, $\IAC = \IDC =0.02\,\Ms$, $\ep = \ez + \ex$, $\omega = \gamma \Hx$, and $T = 300K$.}\label{fig:trajectory}
\end{figure}

In this paper we present a theoretical description of switching using AC/DC spin-current pulses with arbitrary free layer anisotropy and spin-current polarization. Using this description we derive four optimization equations which can be used to numerically calculate the AC/DC spin-current protocol which minimizes Joule heat loss (JHL) during the switching process. The approach used is similar to that used in   \REFS\cite{Tretiakov10,Tretiakov12} which look at minimizing Ohmic losses for ST driven domain wall motion in ferromagnetic wires. As a specific example we calculate the optimal AC/DC spin-current protocol, including the optimal AC and DC pulse times, spin-current strengths, and AC frequency, for a free layers with uniaxial anisotropy and spin-current polarized equally along the easy-axis and hard-axis directions. These results are compared to numeric simulations. We also present a general theoretical prescription for the timer dependent AC/DC spin-current protocol which gives the global minimum JHL for a free layer with arbitrary easy-axis and easy-plane anisotropy strengths. Finally we discuss the range of values the optimal spin-current parameters may take for the practical case where the spin-current parameters are each held constant for the duration of the AC and DC pulses.

\section{Model}
\label{sec:model}
To model the magnetic switching we treat the free layer as a single magnetic domain with a constant saturation magnetization $\Ms$ and  magnetization direction specified by a time-dependent unit vector $\bm(t)$. Its motion is described by the Landau-Lifshitz equation with Slonczewski spin torque term \cite{Slonczewski96,Berger96} (LLS)
\begin{align}
\mathbf{\dot{M}}= \GLL + \GGD + \GST,
\label{eq:LLGS}
\end{align}
where
\begin{align}
\label{eq:Landau_Lifshitz}
\GLL &= -\gamma \bM \times \bHeff\,, \\
\label{eq:Gilbert_damping}
\GGD &= -\gamma\,\alpha\,\bm\times\left[\bM \times \bHeff\right]\,,\\
\label{eq:spin-torque}
\GST &= \gamma\,\Is(t)\,\bm \times \left[\ep \times \bM\right] \,,
\end{align}
are the conservative, dissipative and spin torques acting on the free layer respectively. Here $\gamma$ is the gyromagnetic ratio, $\alpha$ is the dimensionless Landau-Lifshitz damping coefficient, $\Is(t)$ is the time dependent strength of the spin-torque which is proportional to the spin-current density and has units of magnetization, and $\ep$ is the spin-current polarization vector such that $\vIs(t) = \Is(t)\,\ep$. Hereafter we refer to $\vIs$ as the spin-current and $\Is$ as the spin-current strength. Thermal Gaussian noise is included as a random contribution $\mathbf{h}(t)$ to the effective field \cite{Brown63}
\begin{align}
\bHeff (\bm,t) = - \nabla_{\bm} E(\bm) + \mathbf{h}(t),
\end{align}
with correlator given by the the fluctuation-dissipation theorem \cite{footnote}. Here $E$ is the magnetostatic energy density of the free layer
\begin{eqnarray}
E(\bm) = \frac{\Ms}{2} \left[ \Hx\left(1 - \left(\bm\cdot\ex\right)^2 \right) + \Hz\left(\bm\cdot\ez\right)^2 \right]\,,
\label{eq:energy}
\end{eqnarray}
where $\ex$ and $\ez$ are the directions of the easy-axis and easy-plane respectively, and $\Hx$ and $\Hz$ are the strengths of the easy-axis and easy-plane anisotropy fields respectively. The hight of the energy barrier between the two easy-axis directions is $\Eb = \Hx\Ms/2$.

The spin-currents $\vIs(t)$ being considered here consist of two pulses: an AC pulse $\vIs\sin(\omega t)$, followed immediately by a DC pulse $\vIs$ as shown in Fig. \ref{fig:trajectory}. These pulses are characterized by six parameters: the AC driving frequency $\omega$, the AC spin-current strength $\IAC$, the duration of the AC pulse $\tAC$, the DC spin-current strength $\IDC$, the duration of the DC pulse $\tDC$, and the polarization vector $\ep$. A sample {\em non-optimal} switching trajectory from such a pulse is shown in Fig. \ref{fig:trajectory}.

\begin{figure}[H]
  \begin{centering}
  \includegraphics[width=8.6cm]{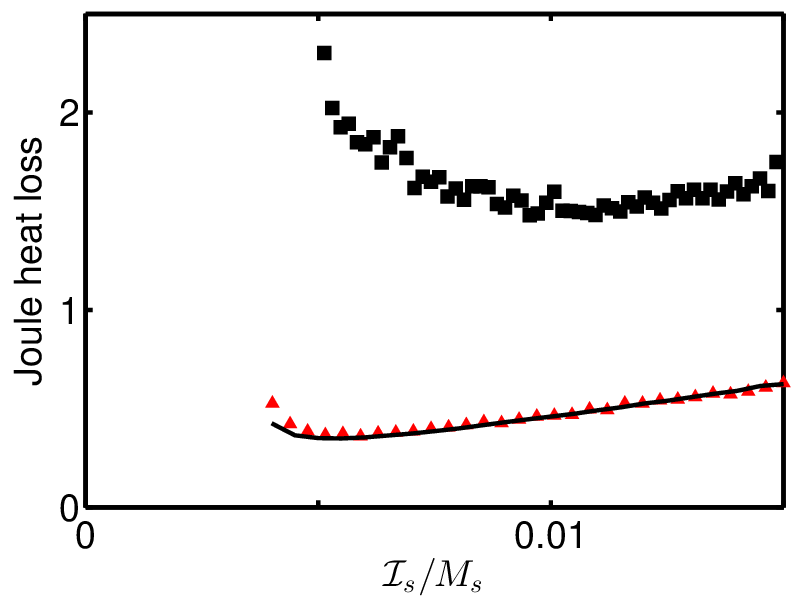}
  \par\end{centering}
  \caption{(Color online) Simulated optimal Joule heat loss in arbitrary units as a function of spin-current amplitude $\Is$ for DC (black squares) and AC/DC (red triangles) spin current methods. Black line represents the theoretical optimal Joule heat lost calculated by numerically solving \EQS\eqref{eq:Opt_Set} while holding $\Is$ constant for both AC and DC pulses. Other parameters the same as in \FIG \ref{fig:trajectory}. }\label{fig:Joule}
\end{figure}

\section{Numerical Simulations}
\label{sec:simulations}
Using the model outlined in the previous section we performed numerical simulations of spin-torque switching using AC/DC spin-currents acting on a free layer with uniaxial anisotropy where $\Hz = 0$, $\Hx = 0.5\Ms$, $\alpha = 0.015$, $\ep = \ez + \ex$, and $T=300K$. For simplicity the strength of the AC and DC spin-current pulses were taken to be equal $\Is = \IAC = \IDC$ for each trial.

To determine the optimal spin-current protocol numerous trials were simulated for a range of AC/DC parameters. Each trial consisted of of $10^3$ switching attempts. For each attempt the magnetization was allow to relax into a thermal equilibrium state before the spin-current pulse was applied. After the pulse the magnetization then was allowed to relax back into a thermal state to determine if the magnetization had switched. The energy loss due to Joule heating for each attempt was then calculated by integrating the power over the duration of the current pulse $J = \int_0^\tpl dt R(\bm)\,\Is^2$. Here $\tpl = \tAC + \tDC$ is the total pulse time and $R(\bm)$ is a resistance that depends on the direction of the magnetization relative to the pinned layer magnetization. In practice this resistance varies little through the switching process \cite{Krivorotov05,Cui10} thus we approximate the JHL as
\begin{align}
\label{eq:JHL_approx}
J = \int_0^\tAC dt R\, \IAC^2\sin^2(\omega t) +  \int_\tAC^\tpl dt R\, \IDC^2\,.
\end{align}
The parameter set resulting in the least JHL, with switching probability above $99.5\%$, for each spin-current amplitude $\Is$ is shown in Fig. \ref{fig:Joule}. Notice the optimal energy loss using this AC/DC method is almost a third of the purely DC method for the same device and the optimal current is roughly half.

\section{AC and DC spin-torque dynamics}
\label{sec:ACDC_dynamics}
In this section we present a theoretical description of free layer magnetization dynamics under the effects of AC and DC spin torque in the absence of thermal noise. The effects of DC spin-current is covered first followed by the effects of AC spin-current.

In order for any spin-current pulse to switch the direction of the free layer the ST must first push the magnetization into an excited state with energy above the height of the energy barrier $E_b$. In this process the ST is opposed by damping which pulls the magnetization to lower energy states. Since the damping torque $\GGD$ always points perpendicularly to the lines of constant energy, known as Stoner-Wohlfarth (SW) orbits, a natural way of assessing the strength of the spin toque is to decompose it into components along the SW orbits and perpendicular to them.

To this end we write the free layer magnetization in terms of the locally orthogonal coordinates $E$ and $\varphi$ such that $\bM = \bM(E,\varphi)$ \cite{Dunn12}.  Here $E$ is nothing more than the energy of the free layer given by \EQ \eqref{eq:energy} and $\varphi = \Omega(E) \tau$ is the time $\tau$ into the SW orbit with energy $E$ in the absence of spin-torque and damping, normalized to $2\pi$. Here $\Omega(E) = 2\pi/\oint d\tau$ is the energy dependent precessional frequency of the magnetization about the easy-axis in the absence of ST and $\tau$ is given by
\begin{align}
d\tau = \frac{\GLL\cdot d\bM}{|\GLL|^2}\,.
\end{align}

The equations of motion for the magnetization in these coordinates are found using the relations
\begin{align}
\dot{E} = -\bHeff \cdot \dot{\bM} \,; \quad\quad
\dot{\varphi} = \Omega(E) \frac{\GLL\cdot\dot{\bM}}{ \left|\GLL\right|^2 }\,.
\label{eq:E_phi_equations_of_motion}
\end{align}
Substituting \EQ \eqref{eq:LLGS} into \EQS \eqref{eq:E_phi_equations_of_motion} for $\dot{\bM}$ the LLS equation becomes
\begin{align}
\dot{E} &= -\alpha U(E,\varphi) + \Is(t)V(E,\varphi)\,, \nonumber\\
\dot{\varphi} &= \Omega(E) - \Is(t)W(E,\varphi)\,.
\label{eq:E_and_varphi_dot}
\end{align}
The three generalized ``forces'' on the RHS of \EQS \eqref{eq:E_and_varphi_dot} represent the effects of dissipation and ST on the free layer energy respectively, as well as the effect of ST on the free layer precessional frequency. They are given by
\begin{align}
U(E,\varphi) &= \frac{1}{\gamma \Ms} \left| \GLL \right|^2 \,;\nonumber\\
V(E,\varphi) &= \frac{1}{\Ms} \left[\GLL\times \bM\right]\cdot\ep \,;\nonumber \\
W(E,\varphi) &= \gamma \Ms \Omega(E) \frac{\GLL\cdot\ep}{ |\GLL|^2 }\,.
\label{eq:Generalized_Forces_E}
\end{align}

For DC spin-currents that are not too large (i.e. on order of the DC critical switching current), the energy $E$ is a slow variable relative to $\varphi$, see Fig. \ref{fig:trajectory}. This separation of time scales allows the equation of motion for $E$ to be averaged over each SW orbit with respect to $\varphi$ \cite{Apalkov05,Dunn11}. Taking this average for DC spin-currents gives
\begin{align}
\dEDC = -\alpha U(E) + \IDC \VDC(E)\,,
\label{eq:E_dot_DC_ave}
\end{align}
where the two $\varphi$-averaged generalized forces are
\begin{align}
U(E) &= \frac{\Omega(E)}{2\pi \Ms} \oint \left[ d\bM\times \bHeff \right] \cdot \bM \,; \nonumber\\
\VDC(E) &= \etax\frac{\Omega(E)}{2\pi \Ms} \oint \left[ d\bM\times \bM \right] \cdot \ex \,.
\label{eq:G_forces_DC}
\end{align}
Here  $\etax = \ep\cdot\ex$ is the portion of the spin-current polarized along the free layer easy-axis, $+\ex$ direction. Note the free layer dynamics depends only on the ST coming from the parallel pinned layer. The lack of any dependence on the ST from the perpendicular pinned layer is the result of the perpendicular ST self canceling as the magnetization precesses about the effective field. It is worth noting that using \EQ\ref{eq:E_dot_DC_ave} one may derive an effective potential energy that includes ST, see \REF\cite{Apalkov05}.

\begin{figure}[H]
  \begin{centering}
  \includegraphics[width=8.6cm]{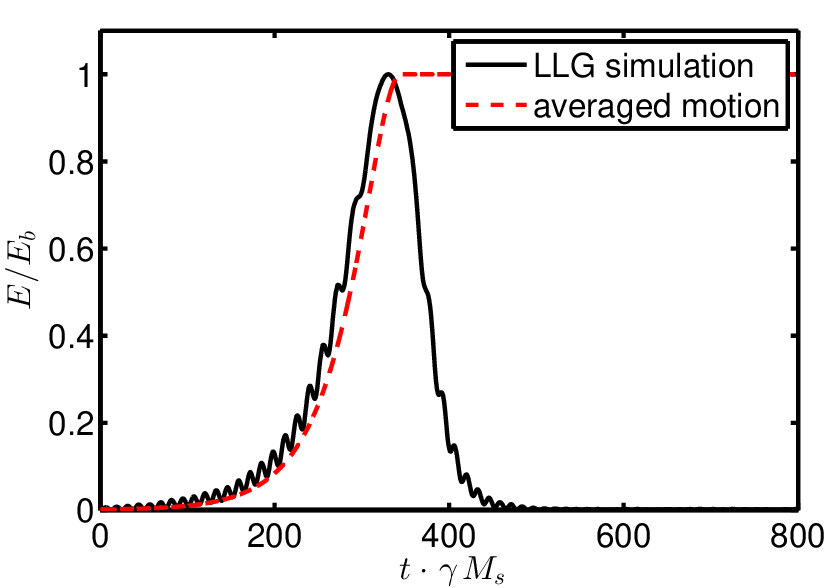}\par
  \end{centering}
  \caption{(Color online) Free layer energy $E(t)$ vs time under the effect of a DC spin-current, calculated via numeric integration of the LLS equation (black,solid) and \EQ \eqref{eq:E_dot_DC_ave} (red,dashed). Here $\Hx = 0.5\Ms$, $\Hz=0$, $\alpha = 0.015$, $\ep = \ex + \ez$, $\IDC=0.02\Ms$, and $T=0$. }
  \label{fig:DC_trajectory}
\end{figure}

A sample DC switching trajectory is shown in Fig. \ref{fig:DC_trajectory} and compared to the $\varphi$ averaged trajectory calculated using \EQ \eqref{eq:E_dot_DC_ave}. Notice the simulated trajectory closely follows the trajectory calculated using \EQ \eqref{eq:E_dot_DC_ave}. Here the small oscillations in the simulated trajectory are the result of the ST from the perpendicular pinned layer and the $\varphi$ precessions which have been averaged over.

\begin{figure}[H]
  \begin{centering}
  \includegraphics[width=7.5cm]{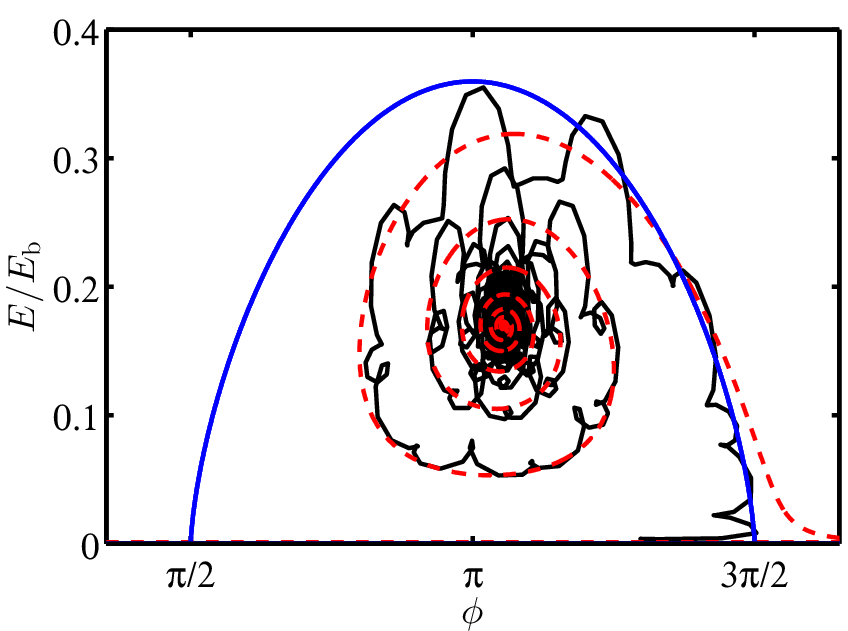}\par
  \end{centering}
  \caption{(Color online) Free layer energy E vs phase $\phi$ under the effect of an AC spin-current calculated via numeric integration of the LLS equation(black,dashed), \EQ \eqref{eq:E_dot_AC_ave} (red,dashed), and $\mathcal{H}=0$ trajectory (blue,solid) from \EQ \eqref{eq:Ham}. Here $\IAC=0.04\Ms$, $\omega = \gamma \Hx$, $\IDC = 0$, and $T=0$ with other parameters the same as in Fig. \ref{fig:trajectory}. }
  \label{fig:AC_trajectory}
\end{figure}

For AC spin-currents with driving frequency $\omega$ close to the natural frequency of the free layer, the free layer magnetization tends to precess with the driving frequency $\omega$. This resonance allows the spin-current polarized along the easy-plane direction to have a non-zero net effect on the energy of the free layer instead of self canceling as in the DC case. The strength and sign of this net perpendicular ST depends on the relative phase $\phi(t) = \varphi(t) - \omega t$ between the magnetization and the AC signal. The equation of motion for this phase is $\dot{\phi} = \dot{\varphi} - \omega$ where $\dot{\varphi}$ is given by \EQ \eqref{eq:E_and_varphi_dot}. Since $\Omega(E) \gg \Omega(E) - \omega$ the phase is also a slow variable relative to $\varphi$. This means the same averaging procedure used in the DC case can be applied to both the $E$ and $\phi$ equations of motion \cite{Dunn12_JAP}. Performing this average gives
\begin{align}
\dEAC &= -\alpha U(E) - \IAC \VAC(E)\sin\phi \,;\nonumber\\
\dot{\phi} &=  \Omega(E) - \omega  - \IAC \WAC(E) \cos\phi \,.
\label{eq:E_dot_AC_ave}
\end{align}
Here the two new generalized $\varphi$-averaged AC forces are given by
\begin{align}
\VAC(E) &= \etaz\frac{\Omega(E)}{2\pi \Ms}\oint \left[ d\bM\times\bM \right] \cdot \ez \cos \varphi \,,\nonumber\\
\WAC(E) &= \gamma\, \etaz\frac{\Omega^2(E)}{2\pi \Ms}\oint \frac{d\bM\cdot\ez}{ |\GLL|^2 }\sin\varphi \,,
\label{eq:G_forces_AC}
\end{align}
and $\etaz = \ep\cdot\ez$ is the portion of the spin-current polarized along the easy-plane direction, $+\ez$-axis. To get the specific form of \EQS \eqref{eq:E_dot_AC_ave} and \eqref{eq:G_forces_AC}, $\varphi = 0$ was chosen to coincide with the easy-plane axis $+\ez$ and the relation $\sin(\omega t) = \sin(\varphi-\phi) = \sin\varphi\cos\phi - \cos\varphi\sin\phi$ was employed. A sample trajectory for $E$ and $\phi$ is shown in Fig. \ref{fig:AC_trajectory} along with the $\varphi$-averaged trajectory calculated via numeric integration of \EQS \eqref{eq:E_dot_AC_ave}. Notice for both the energy first significantly overshoots the equilibrium energy before winding down to it.

\begin{figure}[H]
  \begin{centering}
  \includegraphics[width=8.6cm]{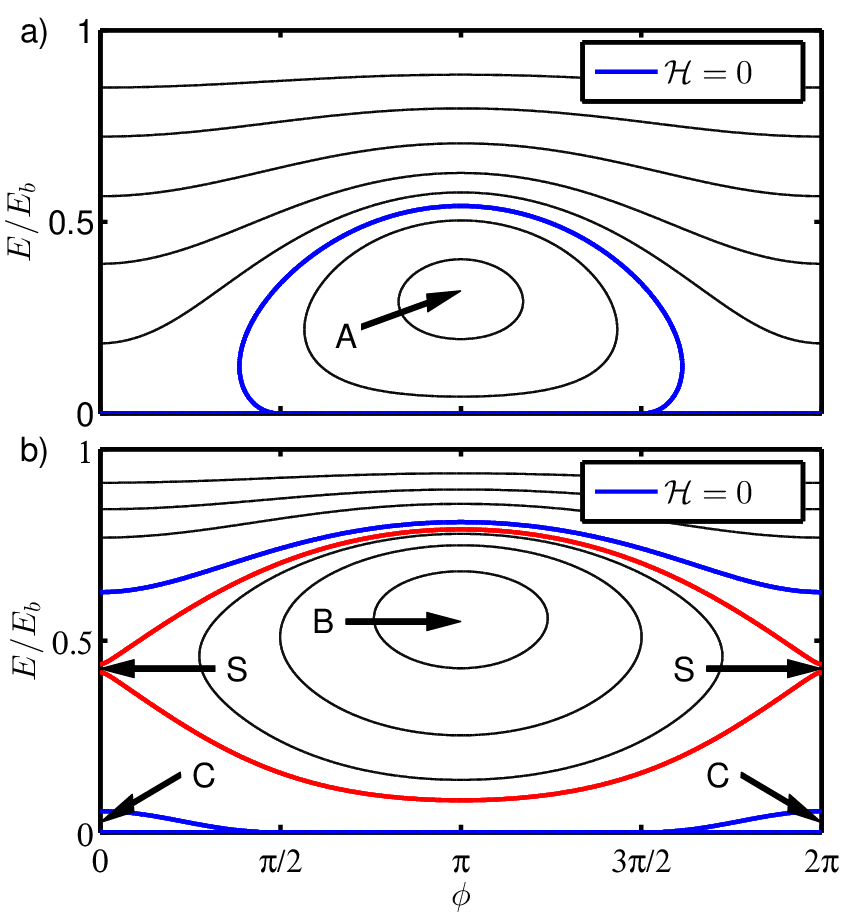}\par
  \end{centering}
  \caption{(Color online) Plots several contours of constant $\mathcal{H}$ (black,thin), including $\mathcal{H}=0$ (blue,thick) and separatrix contour (red,thick) calculated via \EQ \eqref{eq:Ham} for AC frequencies a) $\omega = 0.9\,\Omega_0$ and b) $\omega = 0.7\,\Omega_0$. Stable fixed points are marked as A, B, and C. Unstable fixed point is marked as S. Here $\Hx = 0.5\Ms$, $\Hz=0$, $\alpha = 0.015$, $\ep = \ex + \ez$, and $\IAC=0.04\Ms$. }
  \label{fig:H_contours}
\end{figure}

That the magnetization initial overshoots the equilibrium energy is essential to the efficiency of AC perpendicular spin-currents. To get a better understanding of this energy overshoot one may note that in the absence of damping, $\alpha = 0$, the trajectories given by \EQS \eqref{eq:E_dot_AC_ave} possess an integral of motion and can thus be described as lines of constant value for some function $\mathcal{H}(E,\phi)$ \cite{Dunn12_JAP}. Indeed, one may check that the following function
\begin{align}
\label{eq:Ham}
{\cal H}(E,\phi) = \int\limits_0^E dE'{\cal J}(E') \left[\omega-\Omega(E')  + \IAC\,\WAC(E') \cos\phi \right]  \,,
\end{align}
is conserved by the equations of motion \EQS \eqref{eq:E_dot_AC_ave}, when $\alpha = 0$, if the function $ {\cal J}(E)$ is a solution of the following linear homogeneous differential equation:
\begin{align}
\label{eq:jacobian}
\VAC(E)\frac{d {\cal J}(E)}{dE} = {\cal J}(E)\left(\WAC(E) - \frac{d\VAC(E)}{dE}\right)\,.
\end{align}
Setting $E = 0$ in \EQ \eqref{eq:Ham} gives $\mathcal{H} = 0$, thus for free layer with small initial energies $\Ei \ll \Eb$ under the effect of an AC ST the magnetization should closely follow the $\mathcal{H}=0$ contour for the initial part of its trajectory. Figure \ref{fig:AC_trajectory} shows one such $\mathcal{H}=0$ trajectory along with the simulated trajectory from the LLS equation \eqref{eq:LLGS} and the corresponding $\varphi$-averaged trajectory calculated using \EQS \eqref{eq:E_dot_AC_ave}. Indeed even with damping the magnetization closely follow the $\mathcal{H} = 0$ line for a good portion of its initial upward trajectory.

This means for small damping the dependence of the phase on the energy can be approximated using the $\mathcal{H} = 0$ trajectory as
\begin{align}
\label{eq:AC_cos_phi}
\IAC \cos(\phi(E)) = \mathcal{F}_\Omega(E) - \mathcal{F}_\omega(E,\omega)\,,
\end{align}
where
\begin{align}
\label{eq:AC_F1_F2}
\mathcal{F}_\Omega(E) &= \frac{\int_{0}^E dE' \mathcal{J}(E') \Omega(E')}{\int_{0}^E dE' \mathcal{J}(E') \WAC(E')} \,;\nonumber \\
\mathcal{F}_\omega(E,\omega) &= \frac{\int_{0}^E dE' \mathcal{J}(E') \,\omega}{\int_{0}^E dE' \mathcal{J}(E') \WAC(E')}\,.
\end{align}
Substituting \EQ \eqref{eq:AC_cos_phi} into $\dEAC$ in  \EQ \eqref{eq:E_dot_AC_ave} removes all $\phi$ dependence from the AC energy trajectory and reduces the free layer switching to a one dimensional problem as was the case for DC spin-currents.

\begin{figure}[H]
  \begin{centering}
  \includegraphics[width=8.6cm]{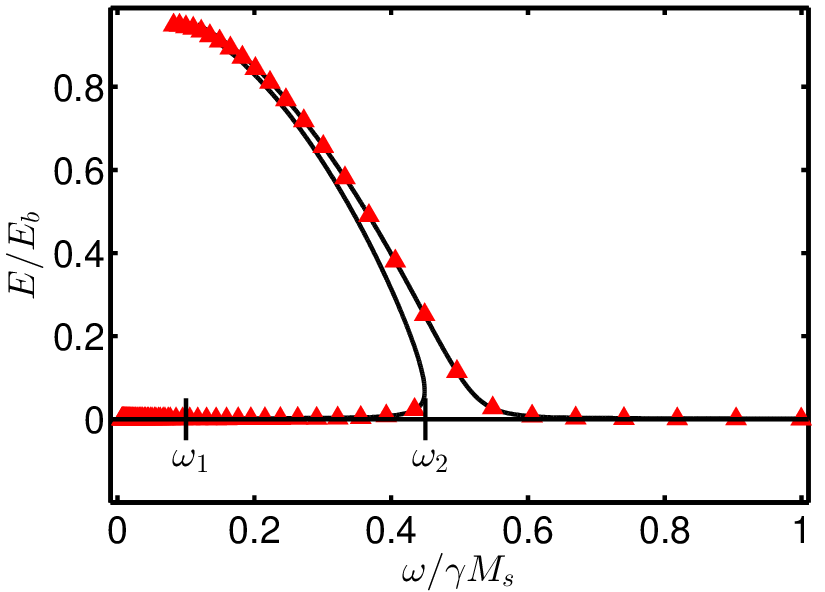}\par
  \end{centering}
  \caption{(Color online) Equilibrium energy as a function of frequency $\omega$  found via simulation of the LLS equation (red,triangles) and calculated numerically from \EQ \eqref{eq:AC_equilibrium_energy}. Same parameters as in Fig. \ref{fig:trajectory}. }
  \label{fig:hyst}
\end{figure}

To end our discussion of AC ST dynamics we now look at the dependence of the energy overshoot on the applied frequency $\omega$. Figure \ref{fig:H_contours} plots several lines of constant $\mathcal{H}$, calculated from \EQ \eqref{eq:Ham}, for a free layer with uniaxial anisotropy and AC signal frequencies $\omega = 0.9\,\Omega_0$ and $\omega = 0.7\,\Omega_0$ where $\Omega_0= \Omega(E=0)$ is the zero energy natural frequency of the free layer. Note the difference in the $\mathcal{H} = 0$ trajectories (blue) as well as the appearance of a separatrix trajectory (red) for $\omega = 0.7\,\Omega_0$. For small initial energies this separatrix prevents the magnetization from reaching the upper $\mathcal{H}=0$ trajectory thus confining it to the lower $\mathcal{H} = 0$ orbit and equilibrium point C. Varying the AC frequency from $\omega = 0.9\,\Omega_0$ to $\omega = 0.7\,\Omega_0$ one sees these two regimes are separated by a bifurcation frequency above which the system has only a single fixed point and below which has two stable fixed points separated by a separatrix.

This bifurcation frequency can be seen clearly in Fig. \ref{fig:hyst}. Here we have calculated the equilibrium energy $\Eeq$ via
\begin{align}
\label{eq:AC_equilibrium_energy}
\IAC^2 = \left(\frac{\alpha U(E)}{\VAC(E)}\right)^2 + \left(\frac{\omega - \Omega(E)}{\WAC(E)}\right)^2\,,
\end{align}
found by setting the LHS of \EQS \eqref{eq:E_dot_AC_ave} to zero, and numerically solving for $E$. Fig. \ref{fig:hyst} also shows $\Eeq$ calculated via simulations of the LLS equation, with temperature $T=0$, by adiabatically varying the frequency with time.
From the simulations we see two {\it jump frequencies} at $\omega_1$ and $\omega_2$ where the energy abruptly jumps between the upper and lower equilibrium branches. Since the initial energy is near zero for any switching process, for $\omega < \omega_2$ the trajectory will overshoot and then relax to the lower equilibrium branch (point C in Fig. \ref{fig:H_contours}) while for $\omega > \omega_2$ the trajectory will overshoot and then relax to the upper (point A in Fig. \ref{fig:H_contours}). As a result $\omega_2 \lesssim \omega$ produces the largest overshoot energies.

\section{Minimizing Joule heat loss}
\label{sec:JHL_minimization}

With a firm understanding of how the free layer magnetization responds to both AC and DC spin-currents we now look for the AC/DC spin-current protocol which minimizes the JHL. We begin by expressing the JHL from a spin-current pulse, given in section \ref{sec:simulations} by \EQ \eqref{eq:JHL_approx}, in terms of the energy/phase coordinates from the previous section. This is done by averaging the power about $\varphi$ resulting in
\begin{align}
\label{eq:JHL_E}
J = \frac{1}{2} \int_\Ei^\Ec dE \frac{R\,\IAC^2 }{\big| \dEAC(E)\big|} + \int_\Ec^\Eb dE \frac{R\,\IDC^2}{\big|\dEDC(E)\big|}\,.
\end{align}
Here $\Ec$ is the energy of the free layer when the spin-current is switched from AC to DC, $\dEAC$ is the AC and DC energy ``velocities'' given by \EQS \eqref{eq:E_dot_DC_ave} and \eqref{eq:E_dot_AC_ave} respectively, and $R$ is an empirical constant proportional to the resistance.  Here also we have eliminated the explicit dependence on time in \EQ \eqref{eq:JHL_E} using the relation $dt = dE/|\dot{E}|$ and each integral represents a path integral over the energy trajectory $E(t)$ of the magnetization during the AC and DC pulses. Using $\Ec$ in place of $\tAC$ and $\tDC$ acts to eliminate one of our spin-current parameters and is centered on the observation that the total pulse time $\tpl$ should be just long enough to produce a switch, thus $\tDC = \tpl-\tAC$ is entirely dependent on $\tAC$ and the remaining spin-current parameters. In this representation the AC and DC pulse times corresponding to $\Ec$ are given by
\begin{align}
\label{eq:tAC_and_tDC}
\tAC =  \int_\Ei^\Ec \frac{dE}{\big| \dEAC(E)\big|}\,;\quad \tDC = \int_\Ec^\Eb  \frac{dE}{\big|\dEDC(E)\big|}\,.
\end{align}
Since the optimal protocol should be such that the free layer is always moving towards higher energy we have drop the modulus signs around $\dEDC$ going forward.


To find the set of AC/DC spin-current parameters which minimizes \EQ \eqref{eq:JHL_E} we take its partial derivatives with respect to each of the remaining spin-current parameters and set them equal to zero. These parameters are: the AC spin-current strength $\IAC$, the AC spin-current frequency $\omega$, the DC spin-current strength $\IDC$, and the energy $\Ec$ where the spin-current changes from an AC pulse to a DC pulse. Taking these derivative, paying careful attention to the dependence of $\phi(E)$ on $\omega$ and $\IAC$, respectively gives
\begin{subequations}\label{eq:Opt_Set}
\begin{align}
 \label{eq:Opt_Set_IAC}
0 &= \frac{\RAC}{2}\int_{\Ei}^{\Ec} dE\, \frac{\IAC}{\dEAC^2} \left( 2\dEAC + \frac{\IAC\,\VAC}{\sin\phi} \right)\,;\\
\label{eq:Opt_Set_omega}
0 &=\frac{\RAC}{2} \int_{E_{0}}^{\Ec} dE\, \frac{\IAC^2}{\dEAC^2} \left(\mathcal{F}_\omega \,\VAC \cot\phi \right)\,; \\
\label{eq:Opt_Set_IDC}
0 &= \RDC\int_{\Ec}^{\Eb} dE\, \frac{\IDC}{\dEDC^2} \Big( \IDC\,\VDC - 2\alpha U \Big)\,;\\
\label{eq:Opt_Set_Ec}
0 &= R\left(\frac{ \IDC^2}{\dEDC} - \frac{ \IAC^2}{\dEAC} \right)_{E=\Ec}\mspace{-36mu}.
\end{align}
\end{subequations}
Here recall the generalized $\varphi$ averaged AC and DC forces are given by \EQS \eqref{eq:G_forces_AC} and \EQS \eqref{eq:G_forces_DC} respectively, $\mathcal{F}_\omega$ is given by \EQ \eqref{eq:AC_F1_F2}, and $\phi(E)$ is given by the $\mathcal{H}=0$ trajectory found using \EQ \eqref{eq:AC_cos_phi}. It should be pointed out that in principle there may be more or fewer optimization equations than those listed in \EQS \eqref{eq:Opt_Set}, the number of which depends on the specific time/energy dependence attributed to each parameter being optimized. For example if one were looking for the optimal protocol with a DC spin current that goes as $\IDC \propto \mathcal{I}_s E + \mathcal{I}_0$ an additional optimization equation, $0 = \partial_{\mathcal{I}_0} J$, would need to be solved in addition to those already listed.

\begin{figure}[H]
  \begin{centering}
  \includegraphics[width=8.6cm]{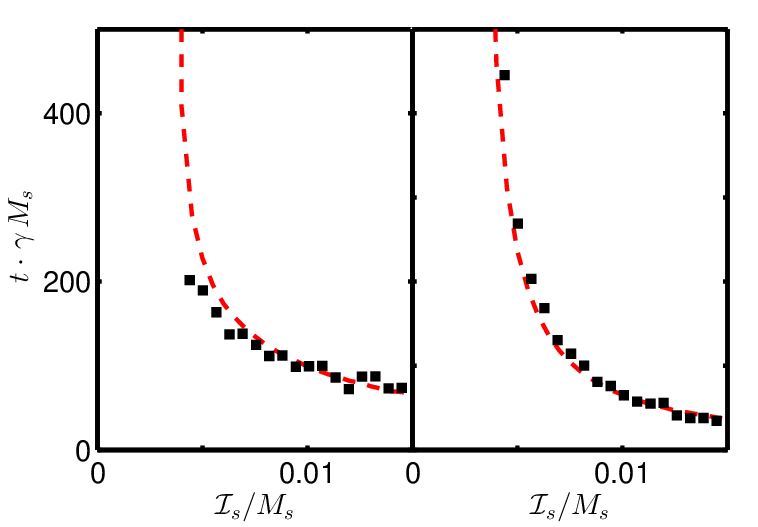}\par
  \end{centering}
  \caption{(Color online) Simulated AC (left) and DC (right) pulse times (black squares) which give the minimum JHL as a function of spin-current amplitude  $\Is$. Red dashed lines represent optimal pulse times numerically calculated using \EQS \eqref{eq:Opt_Set} at each spin-current strength. Same parameters as Fig. \ref{fig:trajectory}. }\label{fig:JHL_sim_vs_num}
\end{figure}

Using these optimization equations the AC/DC spin-current protocol which minimized the JHL can be calculated numerically, with relative ease, for any anisotropy and pinned layer configuration. As an example, we have numerically calculated the optimal spin-current protocol for the free layer simulated in section \ref{sec:simulations} with uniaxial anisotropy and $\ep = \ex + \ez$. The JHL for this protocol is shown in Figs. \ref{fig:Joule} along with the simulated values. The AC and DC pulse times, $\tAC$ and $\tDC$, for these protocols are shown in Fig. \ref{fig:JHL_sim_vs_num} along with the simulated values. Recall, for these simulations $\IAC$, $\IDC$, and $\omega$ were held constant for the duration of each pulse and the strengths of the AC and DC spin-currents were taken to be equal $\IAC = \IDC$.  The first of these restriction defines how $\IAC$, $\IDC$, and $\omega$ behave under integration while the second combines \EQS \eqref{eq:Opt_Set_IAC} and \eqref{eq:Opt_Set_IDC}.

Of course, solving \EQS \eqref{eq:Opt_Set_IDC} numerically, given some assumed form for each parameter, represents at best a local minima in an otherwise infinite parameter space. To find the spin-current protocols which gives the global minimum JHL for the AC/DC spin-current strategies discussed here \EQS \eqref{eq:Opt_Set} must be solved without placing any restrictions on the form of $\IAC$, $\omega$ and $\IDC$. Amazingly, and through no small amount of luck, such a solution is found for \EQS \eqref{eq:Opt_Set}.

To find the spin-current protocol which gives the global minimum JHL for AC/DC spin-current strategies we look for solutions to \EQS \eqref{eq:Opt_Set_IAC}, \eqref{eq:Opt_Set_omega}, and \eqref{eq:Opt_Set_IDC} which make the terms inside the integrals on the RHS identically zero for all values of $E$. One may indeed verify the AC spin-current strength and frequency given by
\begin{align}
\label{eq:Opt_AC}
\IACo(E) = 2\frac{\alpha U(E)}{\VAC(E)}\,, \quad \omo(E) = \Omega(E)\,,
\end{align}
satisfies this requirement for \EQS \eqref{eq:Opt_Set_IAC} and \eqref{eq:Opt_Set_omega}, as does the DC spin-current strength
\begin{align}
\label{eq:Opt_DC}
\IDCo = 2\frac{\alpha U(E)}{\VDC(E)}\,,
\end{align}
for \EQ \eqref{eq:Opt_Set_IDC}. This means the optimal AC/DC spin-current protocol is now entirely determined by $\Ec$ which may be found using \EQ \eqref{eq:Opt_Set_Ec}. It is worth pointing out here that had we included in $R$ any dependencies on the free layer energy ,i.e., the angle between the free layer and the reference layer, the optimal AC and DC spin-current protocols given by \EQS\ref{eq:Opt_AC} and \ref{eq:Opt_DC} would still satisfy \EQS\ref{eq:Opt_Set} and therefore minimize the JHL.

\begin{figure}[H]
  \begin{centering}
  \includegraphics[width=8.6cm]{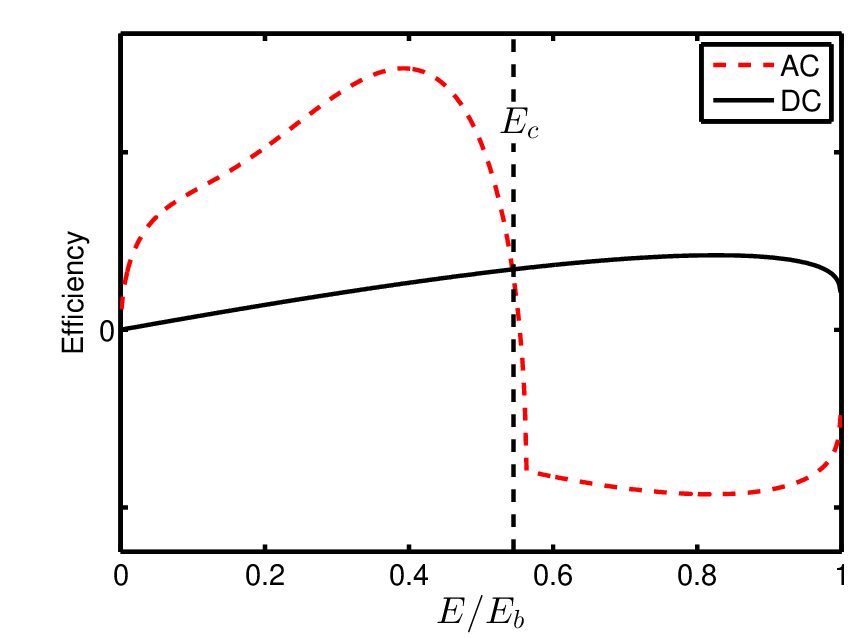}
  \par\end{centering}
  \caption{(Color online) Plots efficiency \EQ \eqref{eq:Opt_Ec_Efficiency} for AC (red,dashed) and DC(black,solid) spin currents for parameters $\Hz = \Ms$, $\Hx = 0.5\Ms$, $\alpha = 0.015$, $\IAC=\IDC=0.04\Ms$ and $\omega = \gamma 2 \Hx$.}\label{fig:Ec_Efficiency}
\end{figure}

One may of course correctly point out there is no guarantee that a solution to \EQ \eqref{eq:Opt_Set_Ec} exists. This apparent contradiction is mitigated by the observation that the AC and DC protocols given by \EQS \eqref{eq:Opt_AC} and \eqref{eq:Opt_AC} are also the optimal protocols of the purely AC and DC spin-current strategies respectively. For purely AC spin-current strategies the optimization equations are given by \EQS \eqref{eq:Opt_Set_IAC} and \eqref{eq:Opt_Set_omega} with $\Ec \rightarrow \Eb$, which $\IACo$ and $\omo$ satisfy. Alternately, for purely DC spin-current strategies the single optimization equation is \EQ \eqref{eq:Opt_Set_IDC} with $\Ec \rightarrow \Ei$, which $\IDCo$ satisfies. This means if \EQ \eqref{eq:Opt_Set_Ec} doesn't have a solution the optimal AC/DC protocol is to use either a purely AC or purely DC spin-current pulse.

For cases where \EQ \eqref{eq:Opt_Set_Ec} has a solution our intuition tells us the optimal protocol should be such that the AC pulse is used until the efficiency of the DC pulse surpasses it thus $\Ec$ should be the energy where the efficiencies of the AC and DC pulses equal. This efficiency criteria is exactly the physical interpretation of \EQ \eqref{eq:Opt_Set_Ec}. Moving the AC terms in \EQ \eqref{eq:Opt_Set_Ec} to the LHS and inverting, keeping the factors of resistance $R$ on each side, gives
\begin{align}
\label{eq:Opt_Ec_Efficiency}
\frac{\dEAC(\Ec)}{R\,\IAC^2(\Ec)} = \frac{\dEDC(\Ec)}{R\,\IDC^2(\Ec)}\,.
\end{align}
The terms on each side are clearly the instantaneous efficiencies of the AC and DC spin-current methods respectively, i.e. the rate of increase for the free layer energy divided by power dissipated in doing so. An example of this is shown in Fig. \ref{fig:Ec_Efficiency}.

\begin{figure}[H]
  \begin{centering}
  \includegraphics[width=8.6cm]{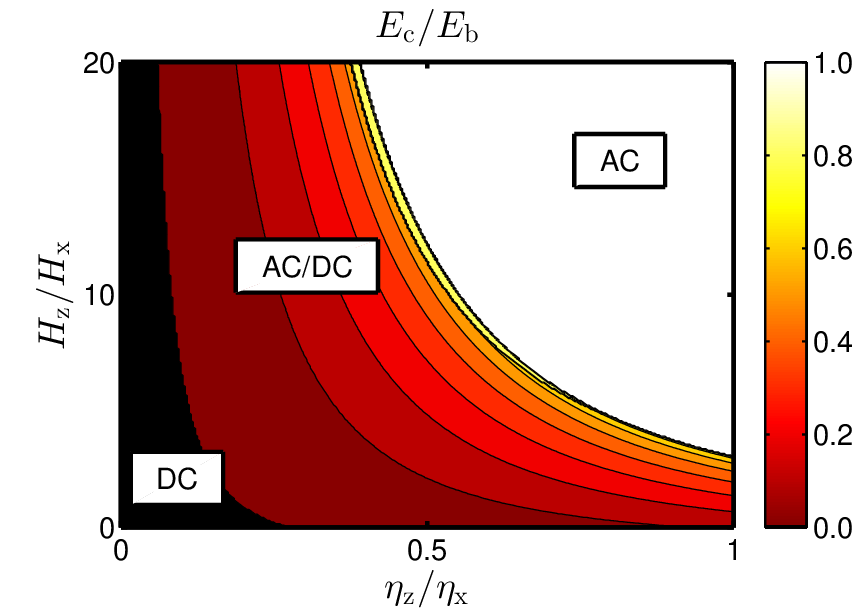}
  \par\end{centering}
  \caption{(Color online) Plots the free layer energy $\Ec$ (in color) where the spin-current should be changed from AC to DC in order to minimize the JHL from switching as a function of the relative anisotropy strengths $\Hz/\Hx$ and the spin-polarization ratio $\etaz/\etax$. Here white indicates the optimal protocol uses only AC spin-current and black indicates the optimal protocol uses only DC spin-current. The AC and DC spin-current protocols are given by \EQS \eqref{eq:Opt_AC} and \eqref{eq:Opt_DC} and $\Ec$ is calculated using \EQ \eqref{eq:Opt_Ec_Efficiency}.}
  \label{fig:ACDC_phase_diagram}
\end{figure}

Whether the optimal AC/DC protocol uses a purely AC, DC, or combined AC/DC protocol naturally  depends on the configuration of the pinned layers and the free layer anisotropy. Fig. \ref{fig:ACDC_phase_diagram} illustrates this dependence by calculating $\Ec$ numerically for a wide range of spin-current polarization and anisotropy configurations using the AC and DC spin-current protocols given by \EQS \eqref{eq:Opt_AC} and \eqref{eq:Opt_DC}. As expected, for strong easy-plane anisotropy and for strong spin-polarization along the easy-plane direction purely AC spin-current protocols are optimal, while for only very weak spin-polarization along the easy-plane direction purely DC spin-current protocols are optimal. This gives a large range of spin-polarizations and anisotropy configurations where AC/DC spin-currents may show improvement over purely AC or DC ones.

To conclude this sections we now discuss the physical significance of the optimal AC and DC spin-current protocols given by \EQS \eqref{eq:Opt_AC} and \eqref{eq:Opt_DC}. From \EQ \eqref{eq:Opt_AC} the optimal AC frequency protocol is clearly to drive the AC spin-current at the energy dependent natural frequency of the free layer $\Omega(E)$. Substituting $\Omega(E)$ for $\omega$ in the equation of motion for $\phi$  \EQ \eqref{eq:E_dot_AC_ave} gives
\begin{align}
\label{eq:dpACo}
\dot{\phi} =  - \IAC \WAC(E) \cos\phi\,.
\end{align}
By inspection one can see \EQ \eqref{eq:dpACo} has one stable equilibrium point at $\phi = 3\pi/2$ and one unstable equilibrium point at $\phi = \pi/2$. This means, by keeping the AC signal frequency at the natural  frequency of the free layer, the magnetization and the AC signal become phase locked. For small initial energies $\Ei \ll \Eb$ this phase locking happens very fast as $\WAC \propto 1/\sqrt{E}$ for $E \ll \Eb$. This phase locking between the magnetization and the AC signal has a profound effect on the efficiency of the AC ST. Placing $\omega = \Omega(E)$ into the AC energy equation of motion \EQ \eqref{eq:E_dot_AC_ave} gives
\begin{align}
\label{eq:E_dot_AC_optimal}
\dEAC = -\alpha U(E) + \IAC \VAC(E)\,,
\end{align}
thus $\omo = \Omega(E)$ also maximizes the ability of the AC ST to push the free layer to higher energy.

To understand the physical significance of the optimal AC and DC spin-current strengths, $\IACo$ and $\IDCo$, one may notice that on applying $\omo$ the AC $\varphi$-averaged energy equation of motion, \EQ \eqref{eq:E_dot_AC_optimal}, takes the exact same form as the DC $\varphi$-averaged energy equation of motion \EQ \eqref{eq:E_dot_DC_ave}. Inserting $\IACo$ into \EQ \eqref{eq:E_dot_AC_optimal} and $\IDCo$ into \EQ \eqref{eq:E_dot_DC_ave} one finds
\begin{align}
\label{eq:E_dot_AC_optimal2}
\dEAC = +\alpha\, U(E)\,.
\end{align}
This is exactly the same as the $\varphi$-averaged AC and DC energy equations of motion {\em without} any spin-current but with time being {\em reversed}. This means for purely AC, purely DC, and for AD/DC spin-current strategies {\em the optimal spin-current protocol exactly time-reverses the purely relaxational trajectory of the free layer magnetization from initial  energy $\Eb$ to the energy minimum $\Ei$}.

One can find further physical significance for the optimal AC and DC spin-currents by noting that for each substituting $\Is = \mathcal{I}_\mathrm{opt}(E)/2$ into the respective energy equation of motion gives $\dot{E} \equiv 0$. This means the optimal AC and DC spin-current strengths are also {\em exactly twice} the local critical currents
\begin{align}
\label{eq:Ic_general}
\IACc(E) = \frac{\alpha U(E)}{\VAC(E)}\,, \quad \IDCc(E) = \frac{\alpha U(E)}{\VDC(E)}\,,
\end{align}
i.e. the spin-current strength needed to perfectly balance damping \cite{Dunn11}.


\section{Discussion}
The spin-current protocol given in the previous section for the global minimum JHL using an AC/DC spin-current strategy provides us with valuable insight into how the free layer should behave near optimal switching conditions. However, applying this strategy is not practical for many reasons. Chief amongst these is its strong dependence on the energy trajectory of the free layer. For ST devices at room temperature thermal noise prevents us from knowing the exact trajectory for each switch. Theoretically these thermal fluctuations could be overcome by adjusting the spin-current in real time.  However, as most practical applications of ST switching require switching times in the nanosecond and sub-nanosecond regimes, any such self adjusting system would be well beyond the limits of current technology. Indeed even without thermal fluctuations producing an AC/DC current which matches the optimal protocol given by \EQS \eqref{eq:Opt_AC} and \eqref{eq:Opt_DC} would likely prove prohibitively difficult. For these reasons, and others not mentioned, we now discuss more practical solutions to \EQS \ref{eq:Opt_Set} using the lessons learned from the previous sections.

The simplest and most practical restriction one can place on the form of the spin-current parameters $\IAC$, $\omega$, and $\IDC$ is to require each be held constant for the duration of their respective pulses, as was done for the simulations presented in section \ref{sec:simulations}. This requires \EQS \ref{eq:Opt_Set} be solved numerically and thus it is useful to place limitations on the range where the optimal values of each parameter may be found.

For constant DC spin-current strengths one finds the optimal DC spin-current must be between twice the smallest and twice the largest local critical current, $2\,\IDCc^\mathrm{min}(E) < \IDCo < 2\,\IDCc^\mathrm{max}(E)$, where $\Ec < E < \Eb$ and $\IDCc(E)$ is given by \EQ \eqref{eq:Ic_general}. These limits are found by noting only the term $\left(\IDC\VDC(E) - 2\alpha U(E)\right)$ in \EQ \eqref{eq:Opt_Set_IDC} may have both negative and positive values which is required for \EQ \eqref{eq:Opt_Set_IDC} to be satisfied upon integration. For most devices the range between these two bounds is relatively small, $2\,\Ic^\mathrm{min}(E) \lesssim 2\,\Ic^\mathrm{max}(E) \lesssim 2\,\Ic$. In such cases the optimal DC spin-current is roughly twice the critical current $\IDCo \simeq 2\,\Ic$ as is observed experimentally for switching using only DC spin-currents. However, for free layers with very weak easy plane anisotropy, i.e. $\Hz/\Hx \lesssim 2$, this range can become quite large. In cases where $\Hz = 0$ the minimum local critical current vanishes, $\Ic^\mathrm{min}(E)\rightarrow 0$ as $E \rightarrow \Eb$. For these devices we note the optimal DC spin-current must also be larger than the DC critical current to ensure switching may occur regardless of our choice for $\Ec$ and in rare cases where the free layer energy decreases due to thermal fluctuations following the AC pulse. This means for free layers with weak easy-plane anisotropy $\Ic < \IDCo < 2\Ic^\mathrm{max}(E)$. For the uniaxial case simulated in section \ref{sec:simulations} we found the the optimal DC current to be slightly larger than the critical current, see Fig. \ref{fig:JHL_sim_vs_num}.

For AC pulses where the frequency is constant one finds the optimal frequency must lie in the range $\omega_2 < \omo < \Omega_0$, where recall $\omega_2$ is the upper bifurcation frequency shown in Fig. \ref{fig:hyst} and $\Omega_0$ is the zero energy natural frequency of the free layer. The lower limit on the optimal AC frequency $\omo > \omega_2$ comes from our desire to maximize the energy overshoot of the free layer. Recall this overshoot is dramatically limited for $\omega < \omega_2$ due to the appearance of a separatrix trajectory which limits the amplitude of the trajectory. The upper limit $\omo < \Omega_0$ is found by noting only the term $\cos\phi$ in \EQ \eqref{eq:Opt_Set_omega} can have both positive and negative values which is required for \EQ \eqref{eq:Opt_Set_omega} to be satisfied upon integration. Here recall $\cos\phi$ comes from the $\mathcal{H} = 0$ AC trajectory and is given by \EQ \eqref{eq:AC_cos_phi}. This behavior can be observed by comparing Figs. \ref{fig:AC_trajectory} and \ref{fig:H_contours}a which have $\omega = \Omega_0$ and $\omega < \Omega_0$ respectively. For the uniaxial case simulated in section \ref{sec:simulations} and for free layers with strong easy-plane \cite{Dunn12} we found $\omo \simeq \omega_2$ is nearly always the case. Of course for real devices $\omega$ may vary slightly between switching attempts thus $\omega$ should be picked such that it balances maximizing the energy overshoot of the free layer with desired switching probability.

For constant AC spin-current strengths establishing a range of possible values for $\IACo$ is significantly more complicated than in the previous two cases. One can formulate upper and lower bounds for $\IACo$ by noting the term $\big( 2\dEAC(E) + \IAC\,\VAC(E)/\sin\phi \big)$ in \EQ \eqref{eq:Opt_Set_IAC} must have positive and negative values in the range $\Ei < E < \Ec$ in order for \EQ \eqref{eq:Opt_Set_IAC} to be satisfied. However, these limits lack simple relationships to measurable physical quantities, such as the critical current and the upper bifurcation frequency, thus we exclude their precise formulation here. Instead, we extrapolate limits from the case where $\IAC$ is held constant and the AC frequency is set equal to the energy dependent optimal value $\omega = \omo(E)$ given by \EQ \eqref{eq:Opt_AC}. For this case \EQ \eqref{eq:Opt_Set_IAC} reduces to  \eqref{eq:Opt_Set_IDC} with $\VDC(E) \rightarrow \VAC(E)$, $\Ec \rightarrow \Ei$, and $\Eb\rightarrow \Ec$. This means the limits on the optimal AC spin-current strength are identical to those in of the DC spin-current with similar substitutions, i.e. the optimal AC spin-current strength must be between the largest local critical current and twice the largest local critical current, $\IACc^\mathrm{max}(E)<\IACo < 2\,\IACc^\mathrm{max}(E)$, in the range $\Ei < E < \Ec$ where $\IACc(E)$ is given by \EQ \eqref{eq:Ic_general}.  Here we have used the alternative lower bounds $\IACc^\mathrm{max}(E)<\IACo$ to insure $\dEAC > 0$ for $E < \Ec$ in light of the observation that, unlike the DC case, as $E\rightarrow0$  $\IACc(E)\rightarrow0$.

Reasserting the restriction that $\omega$ remain constant: for systems where the optimal switching energy is small $\Ec \ll \Eb$ the $\mathcal{H} = 0$ trajectory stays close to $\phi = 3\pi/2$ thus $\sin\phi \lesssim -1$ as is the case in when $\omega = \omo(E)$. This means the limits established previously still apply. From simulations of the free layers in section \ref{sec:simulations} with uniaxial anisotropy we found $\IACo \simeq 1.5\,\IACc(\Ec)$ with $\Ec\simeq 0.25 \Eb$, well within the predicted range. For systems with $\Ec \simeq \Eb$ the AC ST gets weaker as $\phi$ deviates significantly from $3\pi/2$. This corresponds to an effective increase in AC local critical current which in turn shifts the range on the optimal value towards larger currents. Simulations from previous work in Ref. \cite{Dunn12} for purely AC ST switching on a free layer with strong easy-plane anisotropy, $\Hz/\Hx \simeq 30$ and $\ep = \ex + \ez$, show $\IACo \simeq 3 \IACc^\mathrm{max}(E)$ where $\IACc^\mathrm{max}(E)$ is calculated numerically via \EQ \eqref{eq:Ic_general}.

\section{conclusion}
In conclusion we have developed a theoretical description for magnetic switching using consecutive AC and DC spin-current pulses which we have shown may significantly reduce the cost of switching via JHL. In addition we have provided a set of optimization equations which may be used to numerically determine the spin-current protocols which minimize the JHL for such AC/DC given some assumed form for each of the spin-current parameters. This includes determining the optimal AC spin-current strength $\IAC$ and frequency $\omega$, the optimal DC spin-current strength $\IDC$, and optimal AC and DC pulse durations. We have also given the general form of the energy dependent AC/DC spin-current protocol which gives the global minimum JHL for such AC/DC strategies. In all three cases, purely AC, purely DC, and consecutive AC/DC, this globally optimal protocol acts to time reverse the purely relaxational trajectory of the free layer magnetization from the energy barrier to the energy minimum.

\section{acknowledgments}
This work was supported by NSF grant DMR1306734.


\begin{thebibliography}{26}
\bibitem[{foo()}]{footnote}
\bibinfo{note}{Non-equilibrium noise such as spin shot noise is omitted here as
  it is usually weaker than the thermal noise at room temperature
  \cite{Dunn10}}.
\end{thebibliography}
\end{document}